\documentclass[useAMS,usenatbib,letterpaper]{mn2e} 
\usepackage{amsmath}
\usepackage{graphicx}
\graphicspath{{fig/}}

\usepackage{times}

\usepackage[colorlinks,citecolor=blue,linkcolor=magenta]{hyperref}
\usepackage{aas_macros} 
\usepackage{breakurl}

\newcommand{\skipthis}[1]{}
\newcommand{\hii}{{\rm H}{\sc ii}}

\newcommand{\lsim}{${\raisebox{-.9ex}{$\stackrel{\textstyle<}{\sim}$}}$ }
\newcommand{\gsim}{${\raisebox{-.9ex}{$\stackrel{\textstyle>}{\sim}$}}$ }

\def\nh2d{$\rm{NH_2D}$}

\def\nh3{$\rm{NH_3}$}
\def\NH3{$\rm{NH_3}$}

\def\n2hp{$\rm{N_2H^+}$}

\def\h2o{$\rm{H_2O}$}
\def\h2{$\rm{H_2}$}

\def\msun{\,$M_\odot$}

\def\um{\,$\mu\mathrm{m}$}

\def\kms{\,km~s$^{-1}$}
\def\cm2{\,$\rm{cm^{-2}}$}
\def\cm3{\,$\rm{cm^{-3}}$}
\def\cms{\,$\rm{cm^{-2}}$}

\def\vlsr{$v\rm{_{LSR}}$}
\def\vmax{$v_{\rm max}$}
\def\vmin{$v_{\rm min}$}

\def\11{(1,1)}
\def\22{(2,2)}
\def\33{(3,3)}
\def\44{(4,4)}
\def\55{(5,5)}
\def\66{(6,6)}
\def\t21{$T_{21}$}
\def\r31{$R_{31}$}

\def\uv{ $(u,v)$ }

\def\her{\emph{Herschel}}

\let \amp = \&

\title
[Large-scale filaments in Milky Way]
{Large-scale filaments associated with Milky Way spiral arms}
\author
[Ke Wang et al.]
{
Ke Wang, $^{1}$\thanks{E-mail: kwang@eso.org}
Leonardo Testi, $^{1,2,3}$
Adam Ginsburg, $^{1}$
C. Malcolm Walmsley, $^{3,4}$\newauthor
Sergio Molinari $^5$
and
Eugenio Schisano $^5$
\vspace*{6pt} \\
$^1${European Southern Observatory (ESO) Headquarters,
Karl-Schwarzschild-Str. 2,
85748 Garching bei M\"{u}nchen,
Germany}\\
$^2$Excellence Cluster Universe, Boltzmannstr. 2, 85748 Garching bei M\"{u}nchen, Germany\\
$^3$INAF -- Osservatorio astrofisico di Arcetri, Largo E. Fermi 5, 50125 Firenze, Italy\\
$^4$Dublin Institute of Advanced Studies, Fitzwilliam Place 31, Dublin 2, Ireland\\
$^5$Istituto di Astrofisica e Planetologia Spaziali -- IAPS, Istituto Nazionale di Astrofisica -- INAF, via Fosso del Cavaliere 100, 00133 Roma, Italy
}

\begin{document}
\date{Accepted 2015 March 31.  Received 2015 March 31; in original form 2015 February 1}
\maketitle

\begin{abstract}
The ubiquity of filamentary structure at various scales through out the Galaxy has triggered a renewed interest in their formation, evolution, and role in star formation. The largest filaments can reach up to Galactic scale as part of the spiral arm structure. 
However, such large scale filaments are hard to identify systematically due to limitations in identifying methodology (i.e., as extinction features).
We present a new approach to directly search for the largest, coldest, and densest filaments in the Galaxy, making use of sensitive \her\ Hi-GAL data complemented by spectral line cubes. {We} present a sample of {the} 9 most prominent \her\ filaments, {including 6 identified} from a pilot {search} field {plus 3 from outside the field}. These filaments measure 37--99 pc long and 0.6--3.0 pc wide
with masses (0.5--8.3)$\times10^4$\msun, and {beam-averaged} ($28''$, or 0.4--0.7 pc) peak H$_2$ column densities of (1.7--9.3)$\times 10^{22}$\cms. {The} bulk of the filaments are relatively cold (17--21 K), while {some local clumps have a} dust temperature up to 25--47 K. All the filaments are located within {$\lsim$60 pc from the}
Galactic mid-plane.
{Comparing the filaments to a recent spiral arm model incorporating the latest parallax measurements, we find} that 7/9 of them reside within arms, but {most are} close to arm {edges}.
These filaments are comparable in length to {the} Galactic scale height and therefore are not simply part of a grander turbulent cascade.
\end{abstract}


\section{Introduction} \label{sec:intro}

The Milky Way is a barred spiral galaxy. Stars form in dense molecular clouds concentrated on the Galactic disk.
Giant molecular clouds (GMCs), the birth places of high-mass stars, are often organized in complex filamentary networks \citep[e.g.,][]{Bally1987, Busquet2013}. Extremely long and filamentary GMCs can reach up to Galactic scale as part of spiral arm {structures}.
Examples are {the} two most prominent infrared dark clouds (IRDCs):
the ``Nessie'' and ``Snake'' nebulae.
\cite{Goodman2014} present an updated view of the physical Galactic plane and a careful analysis of CO data and find that the 80 pc IR-dark Nessie \citep{Jackson2010} is part of a much longer bone-like structure (up to 5 times the IR-dark extent) tracing the center of the Scutum-Centaurus spiral arm. On the northern part of the same arm lies the Snake, a $>$30 pc long sinuous IR-dark filament, although its molecular extent has not been explored \citep{me14,my-Springer-sum}.
{The} Nessie and Snake both contain a chain of $\sim$1 pc clumps likely fragmented from the pristine filament; some of those massive ($10^2-10^3$\msun) clumps are collapsing and forming high-mass star clusters \citep{Jackson2010, Henning2010, me14, my-Springer-sum}. Thus, Snake- and Nessie-like filaments are directly linked to high-mass star/cluster formation, and bridge local to Galactic structures. These large scale filaments are one to two orders of magnitude larger than the filaments found in the Gould Belt clouds \citep{Andre2013-ppvi}.

If similar large scale filaments are common features of spiral arms, many should be observed towards other {lines-of-sight}.
However, {the} Snake and Nessie are the only two examples reported to be associated with the Milky Way's spiral arm structure. 
Hardly any other similar filaments are known \citep[cf.][]{Ragan2014-GFL} despite {the fact} that the largest molecular cloud complexes are indeed good tracers of spiral structure \citep{Dame1986}.
Moreover, {previous studies associate filaments with spiral arms in the longitude-velocity space alone. A recent spiral arm model has incorporated parallax distance of the arms \citep{Reid2014}, allowing a direct association with important distance information} (\S~\ref{sec:arm}).

\begin{figure*}
\centering
(a)
\includegraphics[width=0.25\textwidth,angle=0]{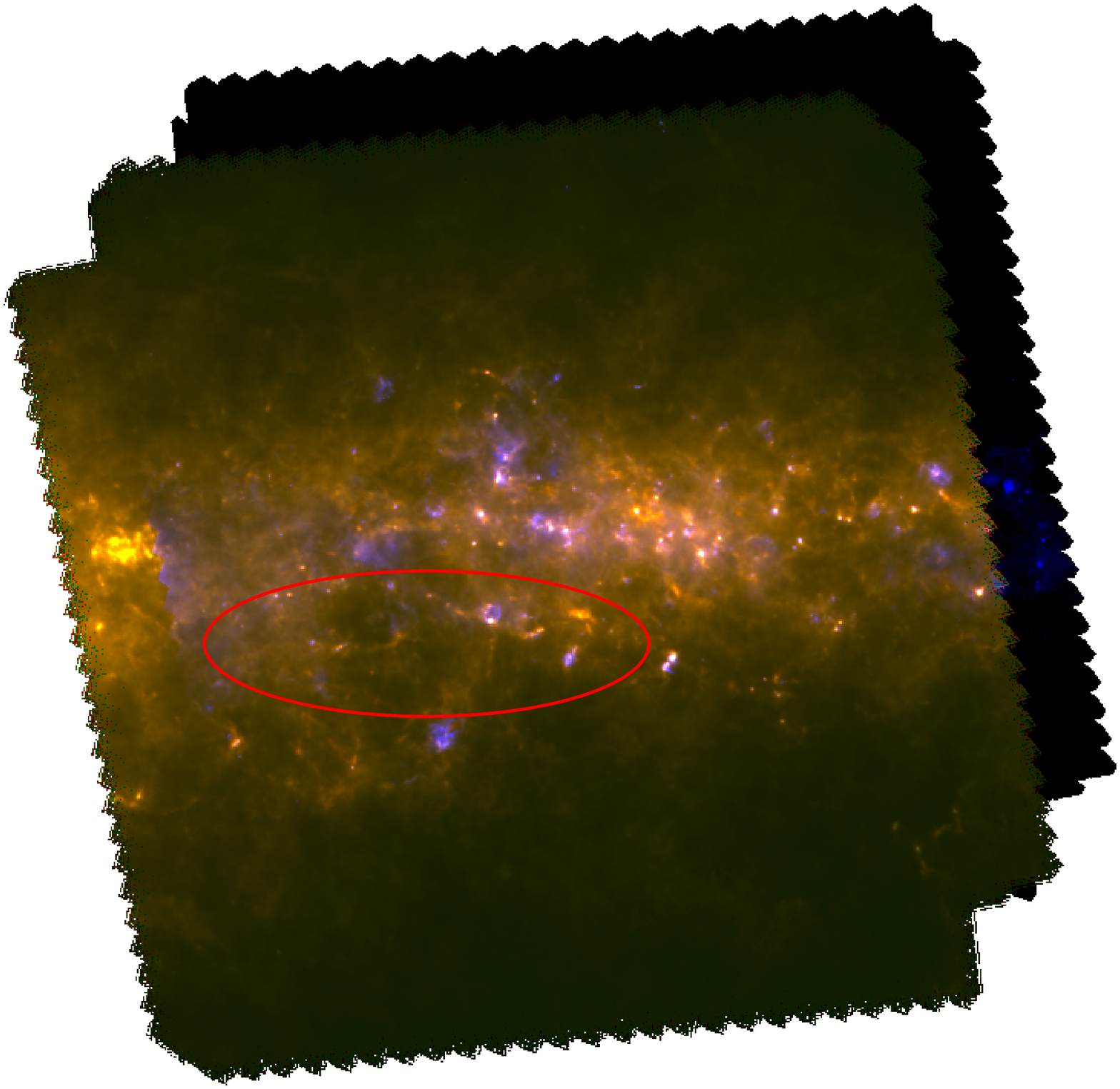}
(b)
\includegraphics[width=0.28\textwidth,angle=0]{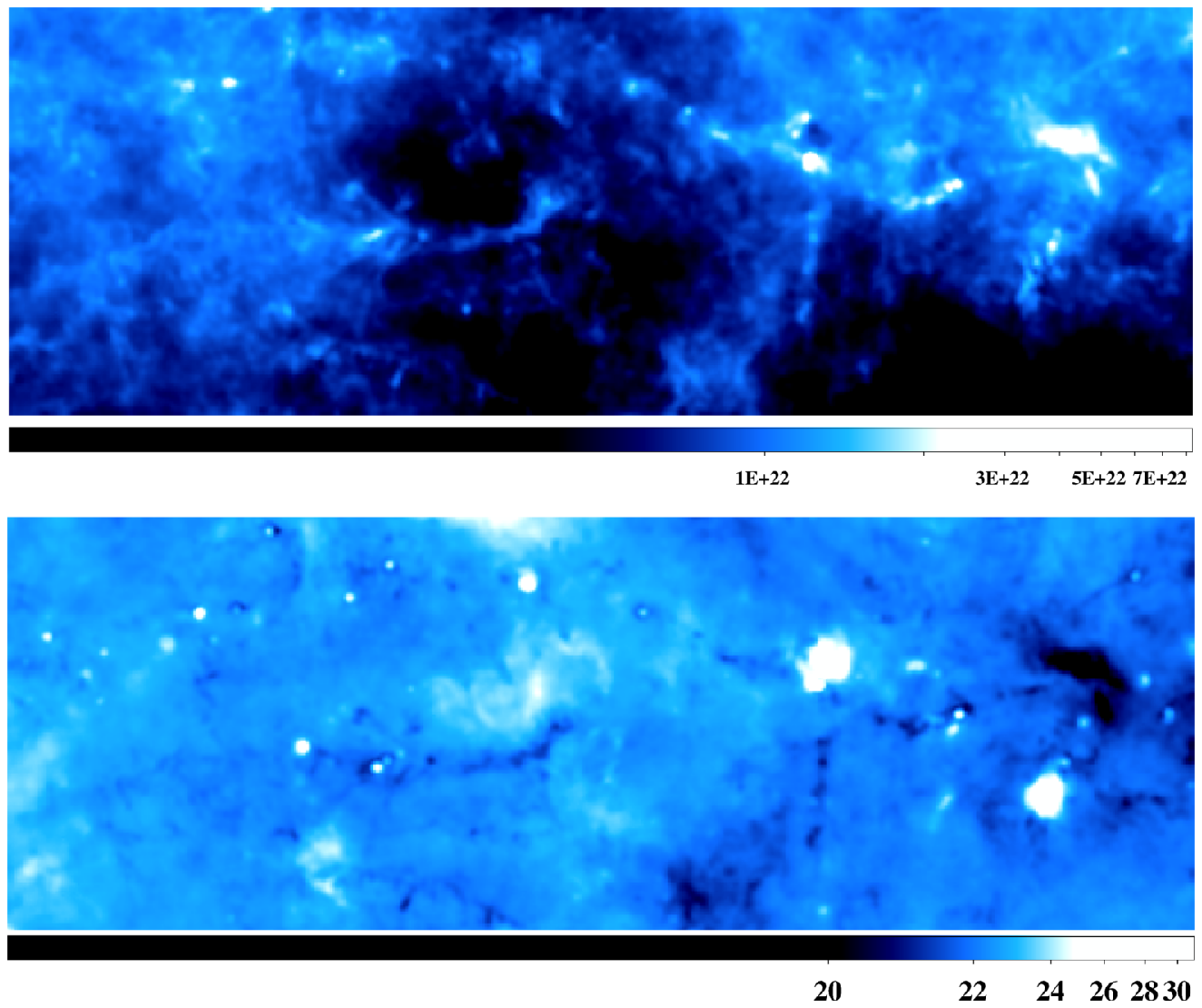}
(c)
\includegraphics[width=0.35\textwidth,angle=0]{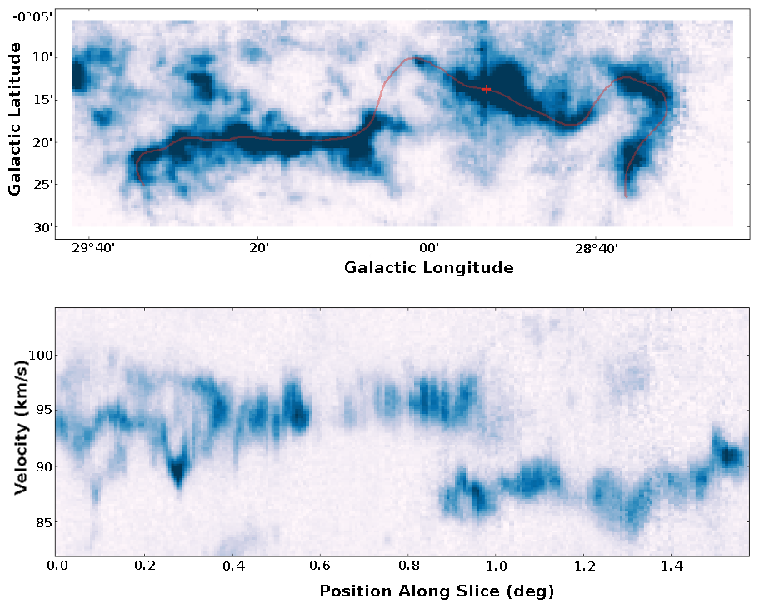}
\caption{Three steps in finding large scale filaments.
(a) Composite image where 500/350/70\um\ images are color coded as red/green/blue. The red ellipse outlines the candidate filaments.
(b) Column density ({\cms,} upper) and dust temperature ({K,} lower) maps.
(c) Customized PV diagram made following the curvature of the filaments. Clearly two filaments (G28, G29) are shown in the lower panel. The red line in the upper panel denotes the PV slice path, and the red cross marks the overlap of the two filaments in 2D.
}
\label{fig:sed}
\end{figure*}

Very long and skinny IRDCs like {the} Snake and Nessie are rare features in IR surveys \citep{Peretto2009} because {(a)} they need a favorable location against background in order to be seen in extinction, {and (b)} they are usually made of individual compact extinction regions that are not always {obviously connected}.
This background dependent method is therefore biased (as reflected by the term ``infrared dark cloud'' itself). See \cite{Wilcock2012} and \cite{Ragan2014-GFL} for more discussion on the limitations in the extinction method.

A more robust way to identify filaments is to directly see them in emission. Since the bulk of the filaments are cold (\lsim 20 K), {they are most readily detected} in far-IR. In this study, we make use of the \her\ Infrared Galactic Plane Survey (Hi-GAL, \citealt{Hi-GAL}) to conduct the first systematic search for Snake- and Nessie-like filaments in the Galaxy. We demonstrate the feasibility of our new identification procedure and present first results from the longitude range of $15^{\circ} < l < 56^{\circ}$. In forthcoming papers we will extend the study to the entire Galactic plane complemented with ongoing spectral line surveys and investigate structure and kinematics of the identified filaments.

\section{Data}
The \her\ key project Hi-GAL \citep{Hi-GAL}, the first unbiased Galactic plane survey at far-IR wavelengths, provides a sensitive data set to \emph{directly} search for large scale, cold and dense filaments. Hi-GAL covers the entire Galactic plane with nominal $|b| \leq 1^{\circ}$ (following the Galactic warp). The data set is comprised of continuum images at 70, 160, 250, 350, and 500\um\ 
obtained with PACS \citep{Poglitsch2010} 
and SPIRE \citep{Griffin2010} 
cameras onboard the \emph{Herschel Space Observatory}
\citep{Pilbratt2010},
with nominal beam sizes of 5.2$''$, 12$''$, 18$''$, 25$''$, and 37$''$, respectively. 
The flux uncertainty in Hi-GAL data is typically 20\%.

Spectral line cubes are used to check the velocity coherence of the filaments and verify that a given 2D filament is indeed a single object, not multiple, unrelated objects along the line of sight.
We retrieve the $^{13}$CO (1--0) data cubes from the Galactic Ring Survey (GRS, \citealt{Jackson2006}). The GRS covers the northern Galactic plane ($18^{\circ}\leq l \leq 55.7^{\circ}, |b| \leq 1^{\circ}$; and $15^{\circ}\leq l < 18^{\circ}, |b| \leq 0.5^{\circ}$), with a resolution of $46''$ and an RMS noise of 0.4 K per 0.2\kms\ channel.

{Our sample also includes three filaments outside the GRS coverage: }the Snake (G11), Nessie (G339), and G64 {(\S\,\ref{sec:sample}). For these filaments we use additional spectral line data}.
We mapped the entire extent of G11 in $^{13}$CO and C$^{18}$O (2--1) using the Atacama Pathfinder Experiment (APEX)
12m telescope. The observations were conducted on 2013 September 4, 7, 8, and 9, in four on-the-fly maps. The {full width at half maximum (FWHM)} beam is $\sim 28''$ and the $1\sigma$ rms is 0.2 K per 0.1\kms\ channel. For G339 we use $^{13}$CO(1--0) from the Three-mm Ultimate Mopra Milky Way Survey (ThrUMMS; \citealt{Nguyen2015})\footnote{See also \burl{http://www.astro.ufl.edu/~peterb/research/thrumms/papers/SubmittedIntro.pdf}}.
For G64, we use the CfA CO (1--0) survey \citep{Dame2001}.
Detailed analysis of the spectral line data will be presented in a forthcoming paper (Wang et al. in preparation).

\section{Methods} \label{sample}
The structure of the interstellar medium is hierarchical and filamentary at multiple spatial scales. Depending on the decomposition method, filaments can be very complex \citep[e.g.][]{Schisano2014}. Here, we focus on the largest, coldest, and densest filaments with simple morphology seen in Hi-GAL images.
We identify cold filaments with three criteria: morphology, temperature, and velocity coherence, as illustrated in Fig. \ref{fig:sed}.
(a) Morphology.
Using the \her\ images of the ``Snake'' and ``Nessie'' as a guide, we have searched for similar features by visual inspection.
A filament is defined by a skinny long feature with aspect ratio $\gg$10 and showing high contrast with respect to its surroundings. As we are looking for the early phases in evolution, we use the long wavelength images at 350 and 500\um\ primarily. 
(b) Temperature.
We then construct temperature and column density maps from pixel-by-pixel SED fitting (Appendix \ref{sec:sed}). Filaments that exhibit systematically lower temperature with respect to {their} surroundings are selected. The morphology is verified using column density maps.
(c) Velocity coherence.
Finally, all filaments are checked for velocity coherence in a customized position-velocity (PV) diagram 
\textit{following the curvature} of the filament. Coherence means continuous, not broken, {emission and} velocity as a function of position along the filament.

{Compared} to the traditional extinction approach, this direct procedure removes geometric ambiguity and bias. One {important} advantage is the temperature information, enabling us to select the cold (thus pristine) {clouds}, ideal for studying the onset of star formation.
For each selected filament, we remove the Galactic scale background emission from {the} \her\ images using a Fourier Transform based routine (Appendix \ref{sec:bg}). We then fit {the spectral energy distribution (SED)} to the multi-wavelength images to obtain temperature and column density {maps} (Appendix \ref{sec:sed}). The codes are made publicly available (see Appendixes).

\begin{table*}
\centering
\begin{minipage}{178mm}
\caption{Physical Parameters of the Selected Sample \label{tab:sample}}
\begin{tabular}{lcc ccc ccc rcc}
\hline
(1) &(2) &(3) &(4) &(5) &(6) &(7) &(8) &(9) &(10) &(11) &(12) \\
ID &Name	&$v_{\rm min}, v_{\rm LSR}, v_{\rm max}$ &$\sigma_v$	&Distance	&Size	&$T_{\rm dust}$	&mean/max $N_{\rm H_2}$	&Mass 	&$z$	&Class &Arm	\\
	 &	 	&(\kms)&(\kms)	&(kpc)	&(pc$\times$pc)		&(K)	&($10^{22}\,\rm{cm}^{-2}$)			&($10^4$\msun)	&(pc)  \\
\hline
G11	  &CFG011.11$-$0.12	&[27, 30.5, 33] &1.9	 &$3.49^{+0.32}_{-0.36}$	&37$\times$0.6	 	&17-25	&0.8/2.3	&1.1	&11   &S 	&	\\
G24	  &CFG024.00$+$0.48	&[93, 96.0, 99] &3.1	 &$5.20^{+0.18}_{-0.19}$	&82$\times$1.9		&17-31	&0.5/2.6	&8.1	&59   &C 	&	\\
G26	  &CFG026.38$+$0.79	&[44, 47.7, 50] &1.4	 &$3.13^{+0.24}_{-0.25}$	&61$\times$1.1		&18-25	&0.3/2.1	&1.9	&62   &X 	&Scu.\\
G28	  &CFG028.68$-$0.28	&[84, 88.2, 92] &3.0	 &$4.89^{+0.22}_{-0.22}$	&60$\times$1.6	 	&17-32	&0.8/7.6	&5.5	&-8   &S 	&Scu.\\
G29	  &CFG029.18$-$0.34	&[89, 93.8, 98] &3.5	 &$5.15^{+0.24}_{-0.23}$	&99$\times$1.6	 	&20-28	&0.5/2.0	&5.5	&-15  &L 	&Scu.\\
G47	  &CFG047.06$+$0.26	&[53, 57.5, 61] &2.3	 &$4.44^{+0.56}_{-0.56}$	&78$\times$3.0	 	&17-29	&0.2/1.5	&2.0	&39   &C 	&Sag.\\
G49	  &CFG049.21$-$0.34	&[66, 68.5, 72] &1.9	 &$5.41^{+0.31}_{-0.28}$	&85$\times$1.6	 	&21-40	&0.9/7.5	&8.3	&-13  &H,X 	&Sag.\\
G64	  &CFG064.27$-$0.42	&[18, 22.0, 25] &2.0	 &$3.62^{+1.56}_{-1.56}$	&51$\times$2.7	 	&17-35	&0.1/1.7	&0.5	&-3   &L,H 	&Loc.\\
G339  &CFG338.47$-$0.43	&[-43,-37.5, -35] &3.7   &$2.83^{+0.26}_{-0.28}$	&80$\times$1.0	    &17-47	&0.6/9.3	&2.9	&-2   &S,H 	&Scu.\\
\hline
\end{tabular}

\medskip
\textbf{Col. (1)}: ID number used throughout this paper.
\textbf{Col. (2)}: Approximate centroid Galactic coordinates in degrees with heading ``CF'' (cold filament).
\textbf{Col. (3-4)}: Velocities and velocity dispersion determined from $^{13}$CO(1--0), with exception of G11 [$^{13}$CO(2--1), Wang et al. in prep] and G64 [CO (1--0), \cite{Dame2001}].
The \vmin\ and \vmax\ are determined from a PV plot customized following the curvature of the given filament (Fig. \ref{fig:sed}); \vlsr\ and velocity dispersion $\sigma_v$ are determined from Gaussian fitting to a spectrum averaged over the entire filament.
\textbf{Col. (5)}: Kinematic distance calculated using the prescription of \cite{Reid2009} with the latest Galactic parameters from \cite{Reid2014}. For sources with distance ambiguity, near distance is adopted.
The error corresponds to a quoted uncertainty of 5\kms\ in systemic velocity. For G49 we adopt the parallax distance of W51 Main/South \citep{Sato2010-W51}, which is very close to the derived kinematic distance for the filament (5.5 kpc).
\textbf{Col. (6)}: Projected length (following the filament's curvature) and FWHM width (Appendix \ref{sec:profile}), deconvolved with the $28''$ beam size.
\textbf{Col. (7--9)}: Range of dust temperature, mean/peak beam averaged column density, and mass measured from a polygon encompassing the filament.
\textbf{Col. (10)}: Height from the coordinates given in col. (2) to the physical Galactic plane taking into account (a) $z_\odot = 25$ pc and (b) the dynamic center of the Milky Way at Sagittarius A* is $l=359.944^{\circ}, b=-0.046^{\circ}$ \citep{Reid2004,Goodman2014}.
\textbf{Col. (11)}: Morphology class, see \S~\ref{sec:sample} for details.
\textbf{Col. (12)}: Association with spiral arm according to the \cite{Reid2014} model (Fig. \ref{fig:arm}).
\\
\textbf{Individual note}:
G11 --- a well studied IRDC nick named ``Snake'' \citep{me14}. 
G26 --- dense part of a larger, less dense filament reported by \cite{Ragan2014-GFL}.
G28 --- reported by \cite{Schneider2014} but without discussion in the context of Galactic filaments. Associated with IRDC G28.53$-$0.25 \citep{{swift2009}}. 
G49 --- associated with a well studied \hii\ region/GMC complex W51. The filament associated with W51 has been reported previously \citep{Carpenter1998-W51,Parsons2012-W51,Ginsburg2015}. Parameters for G49 are derived for the filament only, excluding the \hii\ region which is at a slightly different \vlsr.
G339 --- IRDC also known as ``Nessie'' \citep{Jackson2010,Goodman2014}.
\end{minipage}
\end{table*}

\section{Results and Discussion} 

\subsection{A sample of the 9 most prominent filaments} \label{sec:sample}

\begin{figure}
\centering
\vskip -0.1in
\includegraphics[width=0.48\textwidth,angle=0]{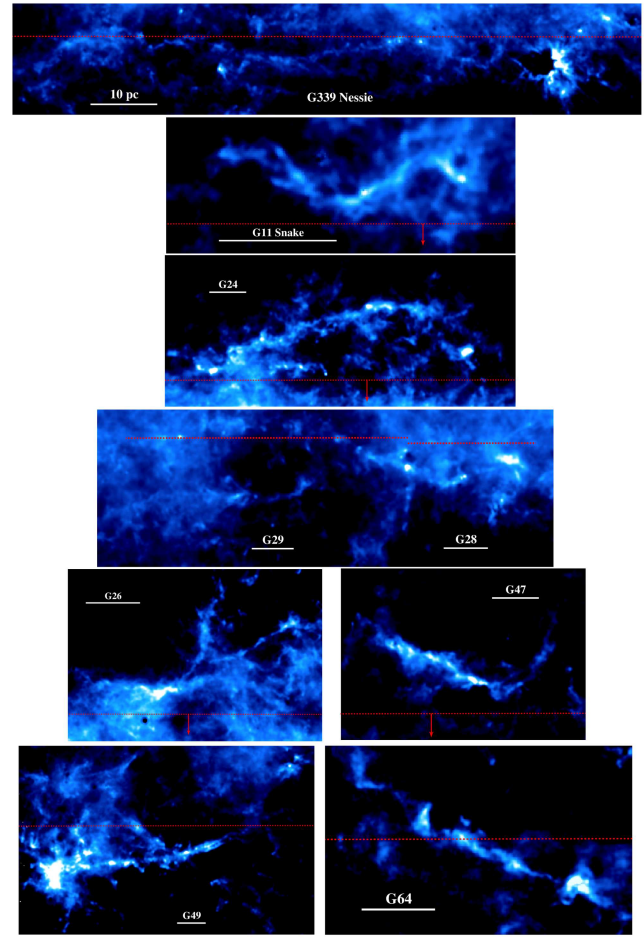}
\caption{
Gallery of the selected filaments: column density maps plotted in logarithmic scale in Galactic coordinates.
{See peak/mean densities for each filament in Table \ref{tab:sample}.}
The scale bars denote 10 pc. {The red dotted line depicts Galactic plane at the source distance;
when accompanied by a red arrow, it means that the plane is outside of the plotted area, parallel to the red dotted line, at the direction of the arrow.
Table \ref{tab:sample} lists distance to Galactic plane for each filament.}
}
\label{fig:gallery}
\end{figure}

We have applied the identification procedure to the GRS field and identified 6 filaments. Outside the GRS field, we include also Snake, Nessie, and G64 {which is identified by chance. They are of the same nature as others and we include them in an effort to compile the most prominent filaments known to date.}
In total, we present 9 filaments (Table \ref{tab:sample}, Fig. \ref{fig:gallery}).
Note that the selected sample is not meant to be complete; instead, it should be regarded as the first descriptive sample of the most prominent, cold and dense, large scale filaments like Nessie and {the} Snake in the Galaxy.
They illustrate the most important visual features and physical characteristics needed to classify GMCs as filaments.

Table \ref{tab:sample} lists physical parameters of the sample. The caption describes how the parameters are derived.
Overall, the filaments measure 37--99 pc long and 0.6--3.0 pc FWHM wide at distances of 2.8--5.4 kpc, with masses (0.5--8.3)$\times10^4$\msun, and beam averaged peak H$_2$ column densities of (1.7--9.3)$\times 10^{22}$\cms. The bulk of the filaments are cold (17--21 K), while {some clumps have higher temperature of 25--47 K most plausibly due to local star formation activities}.
The filaments show remarkable velocity coherence, with small mean velocity gradients of 0.07--0.16 km~s$^{-1}$~pc$^{-1}$ measured along the filaments. In certain clumps, however, the velocity gradients are much larger.
The linear mass densities along the filaments are $10^2-10^3$\msun\ pc$^{-1}$. The maximum linear mass density {that} can be supported by thermal pressure \citep{me11,me14}, given the lowest temperature within each filament, lies in the range of 28--34\msun\ pc$^{-1}$. If the filaments are made of bundles of purely {thermally} supported sub-filaments {like} the Taurus L1495/B213 filament \citep{Hacar2013}, each of these filaments must be comprised of 4--36 thermal sub-filaments.
It is more plausible that they are made of turbulence supported sub-filaments, because the large velocity dispersion of 1.4--3.1 \kms\ suggests the presence of supersonic turbulence spread {throughout} these filaments, similar to high-mass star forming regions and IRDCs \citep{me09,me11,me12,me14}.

We classify the morphology of these filaments {into five categories}:
L: linear straight or L-shape;
C: bent C-shape;
S: quasi-sinusoidal shape;
X: crossing of multiple filaments;
H: head-tail or hub-filament system.
A ``head'' is a prominent/bright clump resides at the tip of a filament.
Some filaments are characterized by more than one class (Table \ref{tab:sample}).
For example, G49 is head-tail (H), and another filament crosses at the middle of the tail (X).
These filaments represent the simplest {morphological modes of filamentary GMCs}.
Different morphology may have resulted from different {filament formation processes} \citep[e.g.,][]{Myers2009a}. Understanding the physics of these simple modes is the first step towards understanding more complex filamentary systems.
While smaller filaments, from sub-pc to 10 pc scales,
naturally occur in turbulent boxes \citep[e.g.][]{Moeckel2014-FLsim}, the filaments we observe are too large to be part of this process.

100 pc appears to be the upper limit of these \her\ identified filaments, whose minimum column density is larger than $\sim$10$^{21}$ \cms. Some, if not all, of them are the dense part of larger, less dense structures. For example, Nessie can be traced up to 430 pc in CO \citep{Goodman2014}. Another example is G26, which is the dense and cold part of a much larger CO filament
\citep{Ragan2014-GFL}. The largest coherent CO filament reported so far is the 500 pc ``wisp'' located at $l \sim 50^\circ$ \citep{LiGX2013-FL500pc}. Those ``CO filaments'' have minimum column densities $\sim$10$^{20}$ \cms\ \citep{Ragan2014-GFL}, an order of magnitude {lower} than our ``\her\ filaments''.

\subsection{Galactic distribution} \label{sec:arm}

\begin{figure}
\centering
\includegraphics[width=0.38\textwidth,angle=0]{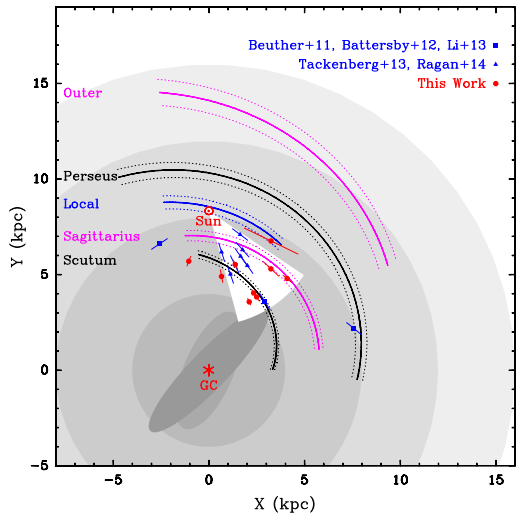}
\caption{
Distribution of the 18 Galactic scale filaments known to date plotted on the latest Milky Way spiral arm model \citep{Reid2014}. {The white circular sector highlights our search field.}
Solid curves are the centers of the best fitted arms and the corresponding dotted lines show {their} widths. Distance error bars for the ``\her\ filaments'' from this work are from Table \ref{tab:sample}; for ``CO filaments'' from the literature \citep{Beuther2011-Coalsack,Battersby2012-FL,LiGX2013-FL500pc,Tackenberg2013,Ragan2014-GFL} a uniform error bar of $\pm0.5$ kpc is plotted for reference.
}
\label{fig:arm}
\end{figure}

All the filaments are located within Galactic mid-plane, with {height ``above/below''} the mid-plane $|z| \lsim 60$ pc.
For comparison, the FWHM thickness of the molecular disk at the Galactocentric radii of these filaments (4--8 kpc) is 80--100 pc ({the} Galactic scale height is FWHM/$\sqrt{8\rm{ln}2} \sim 40$ pc) \citep{Nakanishi2006}, and the intrinsic spiral arm widths are 170--630 pc \citep{Reid2014}. These filaments are comparable {to} or larger than the Galactic scale height and therefore are not simply part of a grander turbulent cascade.

The exact Galactic distribution of these filaments depends on the adopted spiral arm model.
In Fig. \ref{fig:arm} we compare the location of the filaments to the latest spiral arm model of \cite{Reid2014}. The model was derived from VLBA parallax measurements of over 100 masers and represents the most precise model so far. We find that 7 out of the 9 (78\%) \her\ filaments (red symbols in Fig. \ref{fig:arm}) are associated with spiral arms within {their} distance error, whereas {6} of them reside near arm edges and only 1 (G49) close to {an} arm center. Here we have included G339 (Nessie) as ``associated'' if we extrapolate the Scutum arm model (which is limited by observed data points in \citealt{Reid2014}). On the other hand, G11 (Snake) and G24 are outside the $1\sigma$ width of the arms. This is surprising because Nessie and {the} Snake are located right in the center of the Scutum arm when examined in the longitude-velocity plot of the \cite{vallee2008} and \cite{Dame2011} spiral arm models \citep{me14,Goodman2014}.
Expanding the statistics to all the 18 large scale filaments known to date (red and blue symbols in Fig. \ref{fig:arm}), 12 of them are associated with arms (extrapolation of {the} arm model {is} needed for two filaments), whereas only 2 are close to {an} arm center.

If we compare to the \cite{vallee2008} spiral arm model in the traditional fashion, i.e., in $l-v$ space \citep[e.g.][]{Ragan2014-GFL,Goodman2014}, 5/9 of the \her\ filaments, or 8 of the 18 \her\ and CO filaments, are associated with arms. The only filament in our sample that is outside arms in both models is G24 (however it does not stand out in any of the physical parameters listed in Table \ref{tab:sample}, {cf. \citealt{Eden2013}}). 
The inconsistency is due to the difference in the two spiral arm models. Because the \cite{Reid2014} model has accurate distance, it should be regarded as the most up-to-date model.
In summary, {comparing to the Reid et al. model,}
67\% (12/18) of all known large scale filaments are located within spiral arms, whereas 11\% (2/18) reside close to {an} arm center.
{A simple statistical test shows that our \her\ filaments lie preferentially on spiral arms, while the CO filaments identified by \cite{Ragan2014-GFL} is randomly distributed (Appendix \ref{sec:test}).}

How many large scale filaments are there in the Galaxy?
Our systematic search provides {a} rough estimate.
In the GRS field ($l$ runs from 15$^\circ$ to 56$^\circ$) we have identified 6 filaments.
The largest distance is close to $\sim$6 kpc. This search field is a circular sector that occupies 12.88 kpc$^2$ of the Galactic plane area {(Fig.~\ref{fig:arm})}.
If large scale filaments are distributed uniformly in the disk at Galactocentric radii of $R \in [3, 8.5]$ kpc,
we estimate about 90 Nessie-like features. Similarly, \cite{Ragan2014-GFL} identified 7 CO filaments in the GRS field where the most distant one is at $\sim$4 kpc, implying about 240 similar CO filaments within the solar circle.

In the coming years, we plan to expand the search to the entire longitude range using Hi-GAL and ongoing spectral line surveys. The ongoing BeSSeL project \citep{Reid2014} and in the near future, GAIA, will continue to improve the measurement of spiral arms. Spatially resolved simulations have started to provide quantitative constraints on the distribution of large scale filaments with respect to galaxy discs and spiral arms \citep{Smith2014,Dobbs2014}.
Fully operational ALMA and NOEMA will be capable of resolving similar filaments in the LMC, SMC, and nearby galaxies (for reference, 1 arcsec at 10 Mpc corresponds to 48 pc) \citep[e.g.][]{Schinnerer2013-PAWS,Pety2013-PAWS}.
Synergizing {this} information will eventually reveal a full, hierarchical picture of Galactic star formation in the context {of} filamentary structure.

\section*{Acknowledgments} \label{sec:ack}
We {are grateful to} Mark Reid for discussion on Galactic spiral arms and kinematic distance,
{and to}
Phil Myers, Andreas Burkert, Thomas Henning, Henrik Beuther, Sarah Ragan and John Bally for inspiring discussion.
{We appreciate an anonymous referee's valuable comments that helped clarify the manuscript.}
K.W. and A.G. acknowledge support from the ESO fellowship. This study is part of the research project WA3628-1/1 led by K.W. through the DFG priority program 1573 (``Physics of the Interstellar Medium'').

{{\it Herschel} is an ESA space observatory with science instruments provided by European-led Principal Investigator consortia and with important participation from NASA.}
This publication is based on data acquired with the Atacama Pathfinder Experiment (APEX) through ESO program 092.C-0713.
APEX is a collaboration between the Max-Planck-Institut f{\"u}r Radioastronomie, the European Southern Observatory, and the Onsala Space Observatory.
This publication makes use of molecular line data from the Boston University-FCRAO Galactic Ring Survey (GRS). The GRS is a joint project of Boston University and Five College Radio Astronomy Observatory, funded by the National Science Foundation under grants AST-9800334, AST-0098562, \& AST-0100793.
This research made use of open-source Python packages \textsc{Astropy} (\url{astropy.org}), \textsc{glue} (\url{glueviz.org}) and \textsc{pvextractor} (\url{pvextractor.readthedocs.org}).

\bibliographystyle{mn2e_trunc8} 

\newpage
\appendix
\section{FT based background removal} \label{sec:bg}

\begin{figure}
\centering
\includegraphics[width=0.5\textwidth,angle=0]{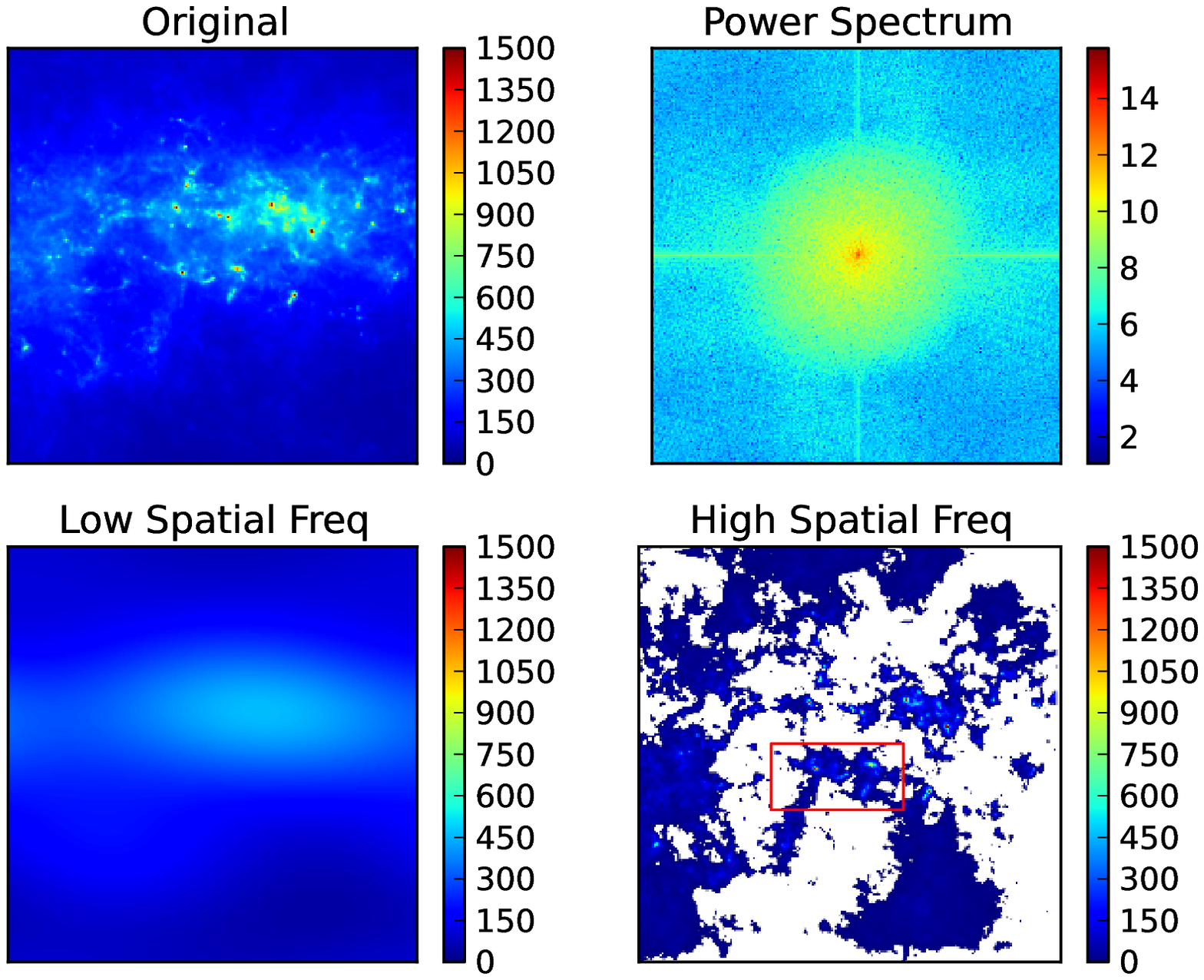}
\caption{
Example of FT based background removal, performed on a $2^{\circ}\times2^{\circ}$ region centered on the filament of interest (red box).}
\label{fig:fft}
\end{figure}

For proper analysis of the structure within the filaments, one has to remove emission from the background and foreground. In the literature, different methods have been used to define background: interpolation across masked-out regions \citep{Peretto2010a}, Gaussian fitting along Galactic latitude \citep{Battersby2011}, or using a constant value from a reference region \citep{Juvela2012}.
These methods involve assumptions {about the structure of the background} or do not {allow for} spatial variation in background. We use a new method based on Fourier Transform (FT). We transform the original image into Fourier domain and separate low and high spatial frequency components, then inverse FT the two components separately back into the image domain. Low spatial frequency component corresponds to large scale background/foreground emission, while high spatial frequency component corresponds to small scale structures.
Fig.~\ref{fig:fft} shows an example in a field approximately $2^{\circ} \times2 ^{\circ}$ centered on the filament G28. One can see that the method separates the Galactic background very well.

The separation between low and high spatial frequencies is determined from the power spectrum. In these images, most power comes from large scale emission, corresponding to low spatial frequencies (the central pixels in power spectrum, equivalent to ``short spacing'' in the \uv\ sampling of interferometric observations). Keeping this monotonic relation in mind, we start with an initial threshold of 90\% of the maximum of the power spectrum.
Pixels above this cut are masked (selected), setting other pixels into zero power, and are inverse Fourier transformed to produce the low spatial frequency map. Similarly, pixels below the cut are transformed to produce the high spatial frequency map.
Images such as Fig.~\ref{fig:fft} are made and inspected, then the cut is changed by a step of $\pm$0.5\%.
This process is iterated until the background image becomes representative of the Galactic scale variation. Our experience shows that $\sim$90\% is a good value for most images. The structure in the resulting background emission has a scale of $\gsim 0.5 ^{\circ}$.

The idea of this method originates from a well-known ``short spacing'' problem in radio interferometric imaging, where the central hole in the \uv\ space caused by a lack of the shortest baselines leads to large scale emission being filtered out. {Similar FT-based filtering} is also widely used {under the name of} ``unsharp masking'' in graphic processing. The method is particularly useful in determining Galactic scale background in \her\ images, but can also be applied to images of any angular size. Compared to other methods, this approach does not involve any assumption {about the underlying structure} and {picks} up spatial variation in background. On top of that, the method is simple and easy to implement. We make the routine publicly available \footnote{\url{https://github.com/esoPanda/FTbg}}.

After removal of background, the derived column density is up to 20--30\% {lower and the} dust temperature is 0.2--2 K lower for the bulk, cold region of the {filament}, than derived from the original images.

\section{SED fitting} \label{sec:sed}

To convert the \her\ continuum images to physical parameters, we have developed a procedure to fit a modified black body function to the multi-wavelength images on a pixel-by-pixel basis \footnote{The SED fitter has been implemented as a standalone python package (\url{http://hi-gal-sed-fitter.readthedocs.org}).}.
In this study, images at 70, 160, 250 and 350\um\ are used while the 500\um\ image is excluded because of {its} low resolution. All images are first convolved to a circular Gaussian beam with FWHM = $28''$ using the kernels provided by \cite{Aniano2011_ker}, and then {re-gridded} to the same pixelization ($8''$ pixel size).
The convolution takes into account the small asymmetry in the \her\ beams.
Pixels below the median value in the 350\um\ images are masked out, and this mask is applied to other wavelengths.
{This mask outlines the approximate boundary of the filament.}
For each unmasked pixel, intensity at various wavelengths are modeled as 
\begin{center}
$I_\nu = B_\nu (1-e^{-\tau_\nu})$,
\end{center}
where
Planck function $B_\nu$ is modified by
optical depth \citep{Kauffmann2008}
\begin{center}
$\tau_\nu = \mu_{\rm H_2} m_{\rm H} \kappa_\nu N_{\rm H_2} / R_{\rm gd}$,
\end{center}
which is caused only by dust.
Here, gas to dust ratio is assumed to be $R_{\rm gd} = 100$. Dust opacity per unit dust mass is \citep{Ossenkopf1994}
\begin{center}
$\kappa_\nu = 4.0(\frac{\nu}{505\, {\rm GHz}})^\beta \rm{cm^{2} g^{-1}}$.
\end{center}
Dust emissivity index is fixed to $\beta = 1.75$ in the fitting.
The free parameters are {the} dust temperature and column density expressed in number of molecular hydrogen {per unit area}. The fitting provides $T_{\rm dust}$ and $N_{\rm H_2}$ maps along with error maps for further analysis.
Typical error in $T_{\rm dust}$ is $<$2 K and $<$20\% in $N_{\rm H_2}$.

\section{Cross-filament profiles} \label{sec:profile}

\begin{figure}
\centering
\includegraphics[width=0.45\textwidth,angle=0]{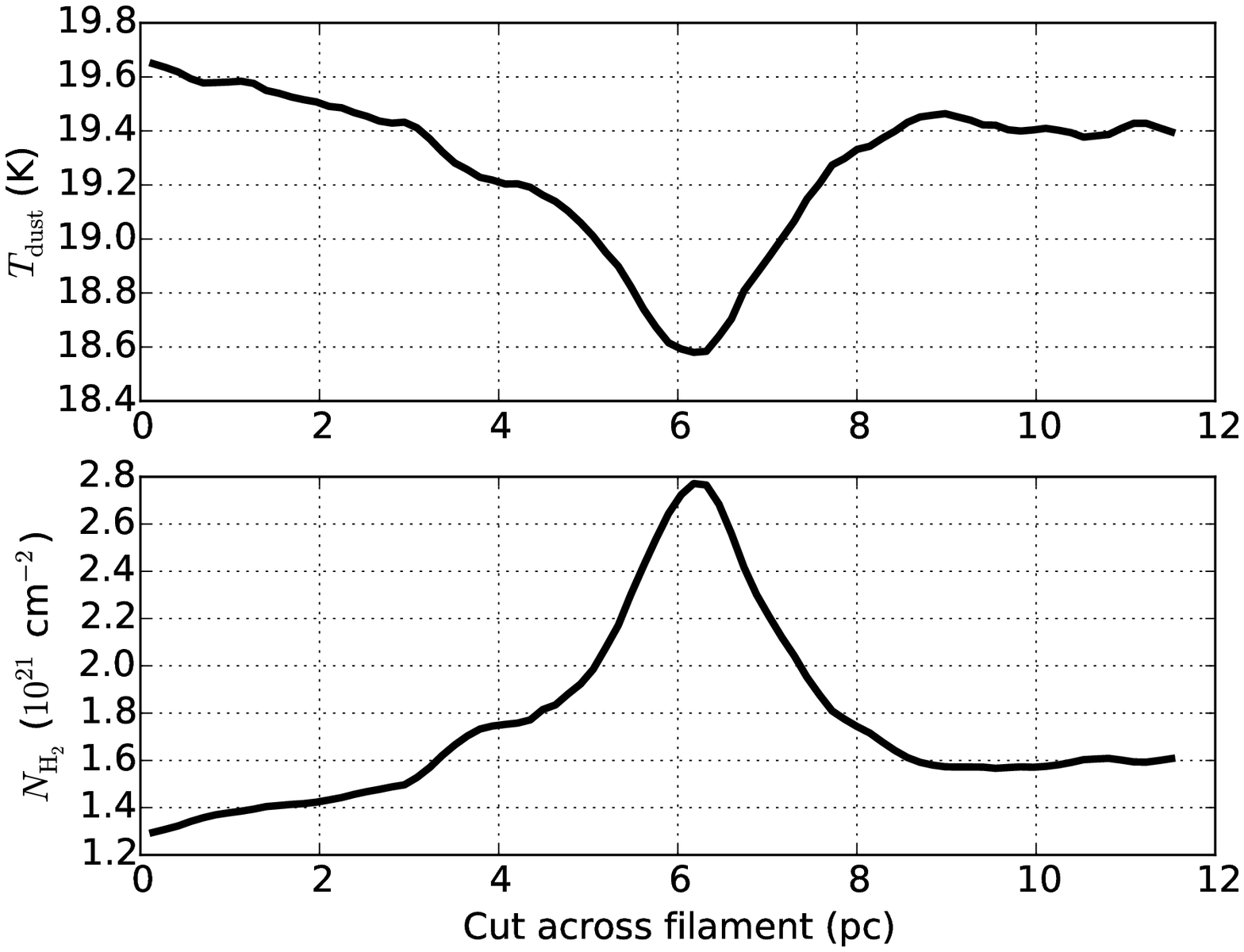}
\caption{
Temperature and density profiles of G64. The cut is perpendicular to the filament; the profiles are averaged within the thickness of the cut ($17.3'$) which is chosen to cover the cold bulk {(the central three clumps)} of the filament.
}
\label{fig:profile}
\end{figure}

Density and temperature profiles are extracted from 1--3 cuts perpendicular to a given filament. The cuts are chosen to be wide enough to represent the general properties of the filament, while narrow enough to avoid curvature. 
{In cases where one cut cannot represent the width of the entire filament, multiple cuts are placed in different representative segments.}
Averaged profiles from the cuts are used for analysis.
Fig.~\ref{fig:profile} shows a representative example.
{As one can see, the}
density and temperature profiles are anti-correlated, i.e., denser regions are in general colder.
The FWHM widths of the filaments are measured {by fitting a Gaussian to the density profile after removal of the ``emission free'' baseline (cut $<$2 and $>$9 pc in Fig.~\ref{fig:profile}),}
and are listed in Table \ref{tab:sample}.
{Uncertainty of the width is 10--30\% depending on how many cuts are averaged.}
{In a forthcoming paper, the profiles will be analyzed with detailed physical models to investigate the structure and stability.}

\section{Statistical test} \label{sec:test}

In the same search field, we identified 6 ``\her\ filaments'' with 5 (83\%) associated with arms, while \cite{Ragan2014-GFL} identified 7 ``CO filaments'' with 3 (43\%) associated with arms.
We regard a filament to be associated with arm if it lies within the width of the arms within its distance error (Fig.~\ref{fig:arm}). Given the large distance error bars, however, the random chance to be associated with arm is relatively high towards some lines of sight.

We perform a simple statistical experiment to test the random probability.
We draw a series of line of sight rays every $2^\circ$ of longitude within the GRS field, and for each ray we calculate the probability of a random filament along this line of sight to fall within a spiral arm. The probability depends on the distance error. For error of 0, $\pm$0.25 (appropriate for our \her\ filaments within the GRS field), and $\pm$0.5 kpc (appropriate for the \citealt{Ragan2014-GFL} CO filaments), the probability averaged for all the lines of sight is 25\%, 37\%, and 48\%, respectively.

Comparing these random probabilities to the observed fractions, the test demonstrates that the \her\ filaments do preferentially lie on spiral arms (83\% vs 37\% random), and the on-arm fraction of the \cite{Ragan2014-GFL} filaments is most likely random (43\% vs 48\% random). This is consistent with the fact that the \her\ filaments have higher average column densities.

%

\end{document}